# Emerging Frontiers: Exploring the Impact of Generative AI Platforms on University Quantitative Finance Examinations

**Rama K. Malladi**, Ph.D., CFA, CAIA, FRM, PMP *

Professor of Finance

Department of Accounting, Finance, and Economics

College of Business and Public Policy

California State University, Dominguez Hills

1000, E Victoria Street, I&I 4320, Carson, CA 90747, USA

Email: rmalladi@csudh.edu        Phone: +1 (310) 243-3560

* Rama K. Malladi is with California State University, Dominguez Hills, California. He received Python and Machine Learning certification in 2021, Chartered Financial Analyst designation in 2006, and Oracle certification in database administration in 1999. This paper benefited from the special "Research with Machine Learning Applications" session at the 2021 Western Economic Association Annual Conference. I would like to thank Bhavna Malladi for validating the results of the three AI Platforms, offering insights into the findings, and editing the paper.

# Emerging Frontiers: Exploring the Impact of Generative AI Platforms on University Quantitative Finance Examinations

**Abstract:**

This study evaluated three Artificial Intelligence (AI) large language model (LLM) enabled platforms—ChatGPT, BARD, and Bing AI—to answer an undergraduate finance exam with 20 quantitative questions across various difficulty levels. ChatGPT scored 30%, outperforming Bing AI, which scored 20%, while Bard lagged behind with a score of 15%. These models faced common challenges, such as inaccurate computations and formula selection. While they are currently insufficient for helping students pass the finance exam, they serve as valuable tools for dedicated learners. Future advancements are expected to overcome these limitations, allowing for improved formula selection and accurate computations and potentially enabling students to score 90% or higher.

## 1. INTRODUCTION

Generative AI[1], or Generative Artificial Intelligence, is an exciting technology field where machines are designed to create original content like text, images, or music (Hu, 2022; Jovanović & Campbell, 2022). It uses complex algorithms to generate content that resembles human creations, pushing the boundaries of what computers can do. One remarkable subset of generative AI is the Large Language Model (LLM), an advanced AI system trained on vast amounts of text and computer code. They possess an impressive ability to understand context, generate coherent responses, and assist in tasks like writing and conversation. LLMs are integral to the generative AI landscape, contributing to its incredible potential and innovation.

OpenAI's ChatGPT[2], Google's Bard[3], and Microsoft's Bing AI (a.k.a. new Bing)[4] are three emerging and popular AI chatbots that rely on Large Language Models (LLMs). Among these LLMs, OpenAI's Generative Pre-trained Transformer (GPT) series has been instrumental in advancing the field. GPT models generate text in different languages and can create human-sounding words, sentences, and paragraphs on almost any topic and writing style—from convincing news articles and essays to conversations in customer-service chatbots or characters in video games. (Brown et al., 2020; Jovanović & Campbell, 2022). Introduced by OpenAI in 2018, GPT models utilize a semi-supervised learning approach, distinguishing them from other prevalent natural language processing models that heavily rely on supervised learning and labeled data.

GPT models revolutionized natural language understanding through generative pre-training and discriminative fine-tuning. Pre-training on various text passages enables them to acquire extensive world knowledge and handle long-range dependencies. This approach empowers GPT models to excel in tasks like question answering, semantic similarity assessment, determination, and text classification using a single task-agnostic model. By combining generative pre-training with discriminative fine-tuning, GPT models have significantly advanced the capabilities of large language models, facilitating more nuanced and accurate natural language processing (Radford et al., 2018).

---

[1] Generative AI Infographic: https://www.visualcapitalist.com/generative-ai-explained-by-ai/

[2] ChatGPT from OpenAI: https://chat.openai.com/

[3] Bard from Google: https://bard.google.com/

[4] Bing AI from Microsoft: https://www.bing.com/search?q=Bing+AI&showconv=1&FORM=hpcodx

AI chatbots in a university setting can be viewed as a double-edged sword. On the one hand, they can gather information from various websites, comprehend context, and provide customized summaries, saving students valuable time that would have otherwise been spent searching the web and clicking on irrelevant links. Regarding academic research, ChatGPT can significantly assist with finance research. There are clear advantages in idea generation and data identification (Dowling & Lucey, 2023).

On the other hand, academic integrity can be severely compromised by AI chatbots. This can be a nightmare for school teachers and university professors as the last thing they may want is grading a submission generated by an AI chatbot but claimed as the original work of a school or university student. Students might find it tempting to use ChatGPT to generate assignment submissions, which would neither help them learn nor develop academically and professionally (AlAfnan et al., 2023).

The primary goal of this study is to examine the capability of AI chatbots in successfully tackling undergraduate-level finance exam questions by delivering accurate and relevant answers. In summary, relying solely on these AI chatbots, what exam score can a student expect to achieve?

In 2018, OpenAI released the first model, GPT-1. In 2019, GPT-2 emerged, showcasing remarkable capabilities in generating text and raising concerns about potential misuse (Solaiman et al., 2019). The year 2020 saw the launch of GPT-3, a significant milestone characterized by its massive size of 175 billion parameters (Brown et al., 2020). GPT-3 is one of the largest publicly-disclosed language models, having been trained on 570 gigabytes of text. By comparison, its predecessor GPT-2, which is functionally similar to GPT-3, had 1.5 billion parameters and was trained on 40 gigabytes of text (Tamkin et al., 2021). The original GPT-1, in contrast, had 0.12 billion parameters. GPT-4 is believed to have more than 1 trillion parameters.[5]

The current iteration of ChatGPT's complimentary version, as of early May 2023, is based on the robust GPT-3.5 architecture. However, it is crucial to note that the subscription-based ChatGPT Plus[6], available at a monthly fee of $20, harnesses the advanced capabilities of GPT-4. Representing a significant leap forward, GPT-4 is a sizable multimodal model capable of processing textual and visual inputs while generating coherent and contextually appropriate textual

---

[5] Comparision of ChatGPT, Bard, and Bing: https://readus247.com/chatgpt-vs-bing-vs-google-bard/

[6] ChatGPT Plus: https://help.openai.com/en/articles/6950777-what-is-chatgpt-plus

outputs. This class of models holds tremendous significance within academic research, as they exhibit substantial potential across a wide range of applications, including but not limited to dialogue systems, text summarization, and machine translation.

By leveraging the power of GPT-4, ChatGPT Plus offers a compelling and engaging experience to its discerning users. GPT-4 is a large multimodal model capable of processing image and text inputs and producing text outputs. Such models are an important area of study as they have the potential to be used in a wide range of applications, such as dialogue systems, text summarization, and machine translation (OpenAI, 2023).

Bard, like ChatGPT, is an LLM that can generate text, translate languages, write different kinds of creative content, and answer your questions in an informative way. However, there are some key differences between the two. The main difference between Bard and ChatGPT is their respective data sources. Bard is trained on an "infiniset" of data chosen to enhance its dialogue and has access to the internet in real-time, whereas ChatGPT is trained on a pre-defined set of data that has not been updated since 2021 (Drapkin, 2023). Bard is trained on a massive dataset of text and code that includes books, articles, code, and other forms of text. ChatGPT is trained on a dataset of text and code that is scraped from the internet. Bard generates creative and imaginative responses that align with storytelling and creative writing prompts. It strives to provide engaging narratives and poetic elements in its responses. ChatGPT, on the other hand, aims to provide informative and contextually appropriate responses across a wider range of topics, including general knowledge, answering questions, and engaging in conversations. Bard uses Google's LaMDA model, often giving less text-heavy responses.

Bing AI starts with the same GPT-4 tech as ChatGPT but goes beyond text and can generate images. Bing adds its Prometheus model on top of GPT-4 (Khan, 2023). Microsoft developed a proprietary Prometheus technology, a first-of-its-kind AI model that combines the fresh and comprehensive Bing index, ranking, and answers results with the creative reasoning capabilities of OpenAI's most-advanced GPT models (Ribas, 2023). While Bing is not primarily an AI chatbot, it incorporates AI and natural language understanding capabilities to enhance search results and provide users with more accurate and contextually relevant information.

One key difference between Bing AI, ChatGPT, and Google Bard is that Bing currently limits conversations to 20 turns, while the other two can continue indefinitely. A turn for Bing AI is a single interaction between a user and the Bing AI chatbot. This interaction can include a user query, Bing AI's response, and any follow-up questions or comments from the user. Bing AI is currently limited to 20 turns per conversation, after which the conversation will end, and context is lost. After reaching a turn limit, any information or context specific to that particular conversation is discarded, and subsequent interactions start with a clean slate. Bing AI is designed to provide factual information and complete tasks, while Bard is designed to have more natural and engaging conversations. So, Bing AI needs to be more efficient with its resources, which is why it has a turn limit. Overall, the 20-turn limit is a trade-off between efficiency and naturalness. The finance exam in this paper has only 20 quantitative questions because of Bing AI's 20-turn limit.

The rationale behind excluding qualitative questions is based on the assumption that qualitative questions are closely linked to quantitative ones. If the AI platforms have successfully answered the quantitative questions, it is expected to have the necessary understanding and context to provide accurate responses to the qualitative questions. Thus, omitting the qualitative questions can help streamline the evaluation process while maintaining confidence in the model's performance.

The concern about academic integrity in university examinations is not new. Students have increasingly turned to online tools to learn and cheat. Homework-helper websites like Chegg can provide valuable support to students, although often, these sites are vulnerable to misuse and represent a significant risk to academic integrity. To effectively address these challenges, universities must understand how homework-helper websites are misused (Pickering & Schuller, 2022).

A wide body of research studies indicates that cheating is rising in higher education (Borgaonkar et al., 2020; Hamilton, 2016; Jr & Keith-Spiegel, 2001; Macfarlane et al., 2014). Business and engineering students have been reported as most likely to cheat (Borgaonkar et al., 2020; Harding et al., 2007; Liu et al., 2015). Research also shows that students who cheat in college are more likely to violate professional ethics when they enter the workforce (Harding et al., 2007). Students now have access to new technological tools, including third-party problem-solving services and extensive online availability of materials (including solution manuals), making it increasingly difficult and challenging to monitor and restrict cheating (Borgaonkar et al., 2020).

The AI platforms add fuel to the fire concerning academic integrity issues. Ethical considerations abound concerning copyright, attribution, plagiarism, and authorship when AI produces academic text. These concerns are especially pertinent because whether the copy is AI generated is currently imperceptible to human readers and anti-plagiarism software (Liebrenz et al., 2023). Some academic publishers and preprints have accepted manuscripts with ChatGPT listed as a "co-author" (Rahimi & Talebi Bezmin Abadi, 2023). Academics have cautioned that It is critical to identify and implement policies to protect against the misuse and abuse of generative AI (Dwivedi et al., 2023).

In March 2023, OpenAI announced the performance of GPT-4 on 34 academic and professional exams. The exams included the SAT, GRE, LSAT, and AP Exams, including AP Microeconomics and Macroeconomics. GPT-4 performed surprisingly well on the exams, scoring in most of the top 10% of test takers. This is a significant achievement, showing that GPT-4 can learn and apply that knowledge in test taking. GPT-4 exhibited human-level performance on most professional and academic exams (OpenAI, 2023). However, Finance is missing from the list of 34 exams.

This paper's objective is not to initiate a tutor prep verification study. This paper includes finance to the list of 34 exams, enabling researchers to calibrate across emerging AI chatbot platforms. This research advances evaluation methods in finance and facilitates cross-platform calibration. The study evaluates the success of undergraduate finance students relying solely on generative AI platforms for complete solutions to exam questions, considering the crucial role of ethics in finance and academic integrity in universities.

The paper is organized as follows: Section 2 describes a typical finance exam paper grouped by difficulty level. Section 3 outlines the results obtained from AI platforms. Finally, in Section 4, the conclusions are summarized, and potential directions for future research are highlighted.

## 2. ASSESSMENT TOOL

The assessment tool has 20 questions (due to Bing AI's limitation of 20 turn limit as described in the previous section). The assessment is divided into five difficulty levels: very easy, easy, medium, hard, and very hard. The assessment is prepared specifically for this paper with new sets of numerical inputs, so the answers are unavailable from previous students posting solutions online. The complete assessment with answers and explanations is in Appendix (A) – Assessment with Solutions and Explanations.

The topics in the assessment are covered in the most commonly used finance textbooks at the undergraduate level (Berk et al., 2015; Brealey et al., 2007; Brigham & Houston, 2021). The covered topics are Business Algebra, Future Value, Present Value, Time Value of Money, Future Value of Annuity, Annuity Payment, Future Value of Growing Annuity, Bond Present Value, Bond Yield to Maturity, Bond Price Change, Net Present Value, Internal Rate of Return, Stock Valuation, Black-Scholes Option Valuation, and Binomial Option Pricing. The assessment questions were not taken from any particular textbook to avoid the possibility of AI platforms prior training from internet sources.

## 3. RESULTS

Credit is granted if the answer provided by the AI platform falls within a range of 99% to 101% of the expected answer, while no credit is awarded for answers outside this range. Appendix (A) contains the precise answers for each set of 20 questions.

### 3.1. ChatGPT from OpenAI

The analysis commences with ChatGPT; the test was executed on May 29, 2023. The condensed results are displayed in Table (1), while a comprehensive feedback report from ChatGPT can be found in Appendix (B) – ChatGPT Results.

At first intriguing, Table (1) unveils the test outcomes of an undergraduate-level finance exam administered to ChatGPT on May 29, 2023. The cumulative exam score is 30%. ChatGPT exhibits commendable proficiency in addressing "Very Easy" questions, attaining a flawless success rate of 100%. However, the model confronts limitations regarding questions categorized as "Hard" or "Very Hard," resulting in 0% correct responses within these challenging levels. Conversely, ChatGPT garners a 25% score for questions under the "Easy" and "Medium" difficulty tiers.

An intriguing discovery unfolds as we delve deeper into ChatGPT's performance. With remarkable acuity, the model accurately discerns the contextual nuances of the questions and comprehends their underlying objectives precisely, achieving a remarkable accuracy rate of 100%. Additionally, ChatGPT demonstrates an impressive ability to independently identify the appropriate formulas to employ, boasting an 80% accuracy.

However, two notable limitations hinder ChatGPT's efficacy in addressing finance questions. Firstly, the model cannot compute exponents and logarithmic functions, an obstacle that impacts its overall performance. Had ChatGPT possessed this capability, it would have achieved a higher score of 75%. Secondly, ChatGPT struggles with iterative computations required for determining solutions to IRR (Internal Rate of Return) and Bond Yield-type questions, hampering its ability to attain optimal results. If the model had the capacity for iterative calculations, its score would have soared to 85%. This multifaceted exploration reveals the strengths and limitations of ChatGPT's performance on the finance exam, shedding light on areas where further improvements could enhance its overall efficacy.

In question 20, one can notice as ChatGPT exhibits signs of generating erroneous information, manifesting as a hallucination. Specifically, the model fabricates a formula for risk-neutral probability, deviating from accurate and established principles. The propensity for hallucinations in AI models poses a significant risk, especially as these models become increasingly convincing and plausible, thereby fostering excessive dependence by users. Paradoxically, the danger of hallucinations intensifies as models exhibit higher levels of accuracy, engendering user trust when providing truthful information in familiar domains. Besides, as these models become integrated into societal frameworks and contribute to the automation of diverse systems, their tendency to hallucinate contributes to the deterioration of overall information quality (OpenAI, 2023).

Equally intriguing is ChatGPT's capacity to recognize its limitations and acknowledge when it encounters insurmountable obstacles. This ability is exemplified in questions 11, 13, and 14, where the model discerns that the problems surpass its current capabilities. Such self-awareness of its limitations is crucial as it motivates the model to strive for improvement and actively learn from these unsolved problems. This adaptive mindset and willingness to tackle challenges are valuable assets for the model's ongoing development and refinement. Finally, it is worth noting that the standard deviation of the discrepancy rate (i.e., ChatGPT answer / expected answer -1) increased as the difficulty level of the questions increased (from 0% to 206%).

| Question | Difficulty | Expected Answer | ChatGPT Answer | Discrepancy | Failure Reason | Strengths | Score |
|---|---|---|---|---|---|---|---|
| 1 | Very Easy | 8.70 | 8.70 | 0% | | Understood the context + Formulated all steps | 5 |
| 2 | Very Easy | 0.0536 | 0.0536 | 0% | | Understood the context + Formulated all steps | 5 |
| 3 | Very Easy | -0.0820 | -0.0820 | 0% | | Understood the context + Formulated all steps | 5 |
| 4 | Very Easy | 0.0972 | 0.0971 | 0% | | Understood the context + Formulated all steps | 5 |
| 5 | Easy | 30,915.47 | 27,091.67 | -12% | Can't compute exponent | Understood the context + Formulated all steps | 0 |
| 6 | Easy | 10,133.46 | 10,135.26 | 0% | | Understood the context + Formulated all steps | 5 |
| 7 | Easy | 0.1012 | 0.0742 | -27% | Can't compute exponent | Understood the context + Formulated all steps | 0 |
| 8 | Easy | 21.51 | 20.32 | -6% | Can't compute log | Understood the context + Formulated all steps | 0 |
| 9 | Medium | 1,363,075 | 4,228,901 | 210% | Can't compute exponent | Understood the context + Formulated all steps | 0 |
| 10 | Medium | 18,505.63 | 23,272.73 | 26% | Can't compute exponent | Understood the context + Formulated all steps | 0 |
| 11 | Medium | 2,236,434 | 0 | -100% | Can't iterate + Missed Formula | Understood the context | 0 |
| 12 | Medium | 1,166.51 | 1,165.06 | 0% | Passed, but missed formula | Understood the context | 5 |
| 13 | Hard | 0.0562 | 0 | -100% | Can't iterate | Understood the context | 0 |
| 14 | Hard | -0.1150 | 0 | -100% | Long numerical computation | Understood the context + Formulated all steps | 0 |
| 15 | Hard | 283.34 | 951.99 | 236% | Can't compute exponent | Understood the context + Formulated all steps | 0 |
| 16 | Hard | 0.1407 | 0.0656 | -53% | Can't compute exponent | Understood the context + Formulated all steps | 0 |
| 17 | Very Hard | 83.93 | 434.35 | 418% | Can't figure out terminal year | Understood the context + Formulated few steps | 0 |
| 18 | Very Hard | 9.7314 | 2.2731 | -77% | Can't compute log | Understood the context + Formulated all steps | 0 |
| 19 | Very Hard | 18.0136 | 4.2772 | -76% | Can't compute log | Understood the context + Formulated all steps | 0 |
| 20 | Very Hard | 6.2388 | 5.58 | -11% | Incorrect formulas (Hallucinate) | Undestood the context | 0 |

| | Standard Deviation of Discrepancy | Grades by Question Difficulty |
|---|---|---|
| Very Easy | 0% | 100% |
| Easy | 10% | 25% |
| Medium | 112% | 25% |
| Hard | 140% | 0% |
| Very Hard | 206% | 0% |
| TOTAL | 126% | 30% |

**Table 1: ChatGPT results**

This table shows the test results of an undergraduate-level finance exam taken by OpenAI's ChatGPT on May 29, 2023.

The cumulative examination score amounts to 30%. ChatGPT demonstrates proficiency in answering all questions categorized as "Very Easy," achieving a 100% success rate. However, it encounters limitations in responding to questions classified as "Hard" or "Very Hard," yielding no correct answers in these levels. Notably, ChatGPT attains a score of 25% for questions falling under the "Easy" and "Medium" difficulty levels.

ChatGPT has figured out each question's context and aim with 100% accuracy. It has developed the right formula to use with 80% accuracy. However, its main drawback in answering finance questions is that it cannot compute exponents and logs – had it done so, it would have received a 75% score. The second drawback is that it cannot compute iteratively to find solutions to IRR and Bond Yield type of questions – had it done so, the score would have been 85%.

## 3.2. Bard from Google

The Bard test run was executed on May 29, 2023. The results are displayed in Table (2), while a comprehensive feedback report from Bard can be found in Appendix (C) – Bard Results.

Compared to ChatGPT, BARD appears comparatively less equipped to address finance-related questions. The cumulative exam score is a mere 15%. Particularly disconcerting is BARD's inability to perform a rudimentary addition operation in response to the first question (Q1). Despite demonstrating accurate comprehension of the contextual goal with a 100% accuracy rate, BARD encountered significant challenges in three critical aspects: delineating an effective pathway to attain the desired goal, discerning the appropriate formulas for each step, and executing precise computations. BARD resorted to employing incorrect formulas from the medium difficulty level and eventually resorted to utilizing hitherto unfamiliar, nonsensical formulas at the hard difficulty level. Compared to ChatGPT, Bard's intermediate steps are cryptic and prone to errors.

As an example, in question 20, Bard says a European call price = S1 * p + S2 * q where S1 = Stock price in the upstate, p = Probability of the stock price going up, S2 = Stock price in the downstate, and q = Probability of the stock price going down. However, no such formula exists, and the real formula looks something like this: $Call\ Value = \frac{(C_u - C_d)(S_0 - S_d/(1+r)^T)}{(u-d)S_0}$, where $C_u$ = call payoff in upstate, $C_d$ = call payoff in downstate, $u$ = stock price in upstate/stock original price, $d$ = stock price in downstate/stock original price, $r$ = risk-free rate, $S_0$ = stock price today and $S_d$ = stock price in a down state.

On a positive note, it exhibited a success rate of 75% when addressing questions classified as "Very Easy." Nonetheless, BARD encountered substantial challenges when confronted with questions of higher difficulty levels, resulting in a notable struggle to provide accurate responses. On selecting appropriate formulas, BARD attained an accuracy of only 55%, accompanied by a 25% occurrence of nonsensical formulas. These outcomes collectively point to a lack of sophistication in BARD's approach, manifested by the absence of intermediate steps that could enhance comprehensibility. Furthermore, BARD's performance was marred by fundamental errors, such as an inability to perform elementary arithmetic operations like addition and division. Notably, Bard, like ChatGPT, makes errors while computing exponents and logarithmic functions.

These findings collectively suggest that BARD's development within the finance domain is still trailing behind ChatGPT's. While BARD frequently produces incorrect answers, it demonstrates a relatively closer approximation to the correct answers than ChatGPT, as evidenced by smaller standard deviations. Specifically, BARD exhibits a lower total standard deviation of discrepancy rate (defined as Bard answer divided by the expected answer, minus one) at 28%, in contrast to ChatGPT's higher discrepancy rate of 126%. Despite its lower scores, Bard appears closer to accurate answers than ChatGPT. One reason could be that "Bard often misrepresents how it works."[7]

When asked why Bard picked more incorrect formulas compared to ChatGPT, the response from Bard was honest and telling. "ChatGPT is also trained on a massive dataset of text and code, but it may be trained on a different dataset than I am. This difference in training data could explain why ChatGPT picks fewer wrong finance formulas than I do. It is also possible that I am not as good at finance as ChatGPT. I am still learning, and I am not perfect. I may make mistakes and not always be able to provide accurate information."

In conclusion, BARD's performance in the finance domain highlights areas that warrant significant improvement, particularly its ability to handle intricate questions, accurately select appropriate formulas and execute fundamental mathematical operations.

---

[7] Bard FAQ: Is Bard able to explain how it works: https://bard.google.com/faq

| Question | Difficulty | Expected Answer | BARD Answer | Discrepancy | Failure Reason | Strengths | Score |
|---|---|---|---|---|---|---|---|
| 1 | Very Easy | 8.70 | 8.80 | 1% | Could not add three prices | Understood the context + Formulated all steps | 0 |
| 2 | Very Easy | 0.0536 | 0.0536 | 0% | | Understood the context + Formulated all steps | 5 |
| 3 | Very Easy | -0.0820 | -0.0820 | 0% | | Understood the context + Formulated all steps | 5 |
| 4 | Very Easy | 0.0972 | 0.0972 | 0% | | Understood the context + Formulated all steps | 5 |
| 5 | Easy | 30,915.47 | 34,728.84 | 12% | Can't compute exponent | Understood the context + Formulated all steps | 0 |
| 6 | Easy | 10,133.46 | 9,245.56 | -9% | Can't compute exponent | Understood the context + Formulated all steps | 0 |
| 7 | Easy | 0.1012 | 0.1115 | 10% | Can't compute exponent | Understood the context + Formulated all steps | 0 |
| 8 | Easy | 21.51 | 22.47 | 4% | Can't compute log | Understood the context + Formulated all steps | 0 |
| 9 | Medium | 1,363,075 | 1,142,330 | -16% | Can't compute exponent | Understood the context + Formulated all steps | 0 |
| 10 | Medium | 18,505.63 | 19,115.88 | 3% | Can't compute exponent | Understood the context + Formulated all steps | 0 |
| 11 | Medium | 2,236,434 | 1,298,129 | -42% | Used a wrong formula | Understood the context | 0 |
| 12 | Medium | 1,166.51 | 1,012.74 | -13% | Used a wrong formula | Did not understand the context | 0 |
| 13 | Hard | 0.0562 | 0.0538 | -4% | Gibberish formula | Did not understand the context | 0 |
| 14 | Hard | -0.1150 | -0.1225 | 7% | Gibberish formula | Did not understand the context | 0 |
| 15 | Hard | 283.34 | 262.05 | -8% | Can't compute exponent | Understood the context + Formulated all steps | 0 |
| 16 | Hard | 0.1407 | 0.1434 | 2% | Gibberish formula | Understood the context | 0 |
| 17 | Very Hard | 83.93 | 154.12 | 84% | Gibberish formula | Did not understand the context | 0 |
| 18 | Very Hard | 9.7314 | 13.65 | 40% | Used a wrong formula | Understood the context | 0 |
| 19 | Very Hard | 18.0136 | 6.35 | -65% | Used a wrong formula | Understood the context | 0 |
| 20 | Very Hard | 6.2388 | 5.31 | -15% | Gibberish formula | Did not understand the context | 0 |

| | Standard Deviation of Discrepancy | Grades by Question Difficulty |
|---|---|---|
| Very Easy | 0% | 75% |
| Easy | 8% | 0% |
| Medium | 16% | 0% |
| Hard | 5% | 0% |
| Very Hard | 56% | 0% |
| TOTAL | 28% | 15% |

**Table 2: Bard results**

The results of Google/Alphabet's Bard on May 29, 2023, are presented in this table. BARD achieved a proficiency level of only 15% in the overall examination. It demonstrated a 75% success rate in answering questions categorized as "Very Easy." However, BARD encountered difficulties with questions at all other higher difficulty levels and struggled to provide accurate responses. When selecting the correct formula, Bard had an accuracy of only 55% and generated nonsensical formulas 25% of the time. Overall, BARD's approach lacked sophistication, as it failed to provide intermediate steps that could aid in understanding.

Moreover, BARD made simple errors, such as being unable to perform basic addition and division operations. These findings indicate that BARD is still lagging behind ChatGPT in terms of its development in the finance domain. While BARD often provides incorrect answers, they are relatively closer to the correct answers than those generated by ChatGPT, as indicated by smaller standard deviations.

### *3.3.  Bing AI from Microsoft*

The Bing AI test run was executed on May 29, 2023. The results are displayed in Table (3), while a comprehensive feedback report from Bard can be found in Appendix (D) – Bing AI Results.

Bing runs on GPT-4, the latest LLM from OpenAI. As OpenAI updates GPT-4 and beyond, Bing benefits from those improvements.[8] Meanwhile, Bard uses Google's own LaMDA LLM. Bing Chat can generate images, while Bard can generate code and explain it. Consequently, it is reasonable to anticipate that Bing would outperform ChatGPT, which utilizes the earlier GPT-3.5 version.

However, disappointingly, this is not the case. Despite Bing's AI performing better than Bard, it does not achieve the same level of proficiency as ChatGPT. In the finance quiz, Bing AI received an overall score of 20%, which is lower than ChatGPT's score of 30%. On the other hand, as readers may recall, Bard scored 15% on the same quiz. Interestingly, Bing AI exhibits a higher standard deviation of discrepancy rate at 37% compared to Bard's rate of 28%. However, this value is significantly lower than ChatGPT's strikingly high rate of 126%.

Like ChatGPT and Bard, Bing AI encounters common challenges, including the inability to correctly compute exponents and logarithmic functions, utilizing incorrect formulas, and introducing a new problem of fake URLs. Bing AI attempts to address one particular issue shared by ChatGPT and Bard: the difficulty in accurately citing and referencing materials. Bing AI provides URLs due to its role as a traditional search engine, employing ranking factors like links and website content. In contrast, ChatGPT lacks internet access and cannot crawl or read web pages, making URL provision impossible. Additionally, while Bing focuses on information retrieval, ChatGPT prioritizes conversational engagement, avoiding disruptions like URL inclusion. When questioned about its inability to cite sources, Bard provides the following response:

"I cannot cite sources because I am a large language model, also known as a conversational AI or chatbot trained to be informative and comprehensive. I am trained on a massive amount of text data, and I can communicate and generate human-like text in response to a wide range of prompts and questions. For example, I can provide summaries of factual topics or create stories.

---

[8] Bing: https://blogs.bing.com/search/march_2023/Confirmed-the-new-Bing-runs-on-OpenAI%E2%80%99s-GPT-4

| Question | Difficulty | Expected Answer | Bing AI | | | | |
|---|---|---|---|---|---|---|---|
| | | | Answer | Discrepancy | Failure Reason | Strengths | Score |
| 1 | Very Easy | 8.70 | 8.70 | 0% | | Understood the context + Formulated all steps | 5 |
| 2 | Very Easy | 0.0536 | 0.0536 | 0% | | Understood the context + Formulated all steps | 5 |
| 3 | Very Easy | -0.0820 | -0.0817 | 0% | | Understood the context + Formulated all steps | 5 |
| 4 | Very Easy | 0.0972 | 0.0972 | 0% | | Understood the context + Formulated all steps | 5 |
| 5 | Easy | 30,915.47 | 32,578.81 | 5% | Can't compute exponent | Understood the context + Formulated all steps | 0 |
| 6 | Easy | 10,133.46 | 9,306.96 | -8% | Can't compute exponent | Understood the context + Formulated all steps | 0 |
| 7 | Easy | 0.1012 | 0.0718 | -29% | Can't compute exponent | Understood the context + Formulated all steps | 0 |
| 8 | Easy | 21.51 | 25.06 | 17% | Can't compute log | Understood the context + Formulated all steps | 0 |
| 9 | Medium | 1,363,075 | 2,039,563 | 50% | Can't compute exponent | Understood the context + Formulated all steps | 0 |
| 10 | Medium | 18,505.63 | 15,096.60 | -18% | Can't compute exponent | Understood the context + Formulated all steps | 0 |
| 11 | Medium | 2,236,434 | 4,107,710 | 84% | Can't iterate + Missed Formula | Understood the context | 0 |
| 12 | Medium | 1,166.51 | 1,048.64 | -10% | Can't compute exponent | Understood the context + Formulated all steps | 0 |
| 13 | Hard | 0.0562 | 0.0516 | -8% | Can't iterate | Understood the context + Formulated all steps | 0 |
| 14 | Hard | -0.1150 | -0.1411 | 23% | Can't compute exponent | Understood the context + Formulated all steps | 0 |
| 15 | Hard | 283.34 | 238.28 | -16% | Can't iterate | Understood the context + Formulated all steps | 0 |
| 16 | Hard | 0.1407 | 0.1208 | -14% | Can't iterate | Understood the context + Formulated all steps | 0 |
| 17 | Very Hard | 83.93 | 27.98 | -67% | Used a wrong formula | Did not understand the context | 0 |
| 18 | Very Hard | 9.7314 | 4.09 | -58% | Can't compute log + wrong formula | Understood the context + Formulated all steps | 0 |
| 19 | Very Hard | 18.0136 | 5.17 | -71% | Can't compute log + wrong formula | Understood the context + Formulated all steps | 0 |
| 20 | Very Hard | 6.2388 | 9.44 | 51% | Can't compute exponent + wrong formula | Understood the context + Formulated all steps | 0 |
| | | | Standard Deviation of Discrepancy | | | Grades by Question Difficulty | |
| | Very Easy | | | 0% | | | 100% |
| | Easy | | | 17% | | | 0% |
| | Medium | | | 42% | | | 0% |
| | Hard | | | 16% | | | 0% |
| | Very Hard | | | 51% | | | 0% |
| | TOTAL | | | 37% | | | 20% |

**Table 3: Bing AI (a.k.a new Bing) results**

The results of Bing AI on May 29, 2023, are presented in this table. Bing AI achieved a proficiency level of only 20% in the overall examination. Bing AI's performance in the exam is assessed at 20%. It displays exceptional proficiency when dealing with questions categorized as "Very Easy," achieving a flawless success rate of 100%. Nevertheless, the model faces limitations when confronted with questions at higher difficulty levels, resulting in a 0% accuracy rate for these challenging tasks. Bing AI exhibits similar flaws to ChatGPT and Bard, including making simple errors like being unable to carry out basic addition and division operations. Bing AI provided URLs for only 45% of the exam questions, mostly "very easy" ones. However, all 35 provided URLs were functional, with 34 highly relevant. In terms of sophistication, accuracy, explanation of steps, and formulas, Bing AI demonstrates a moderate level of proficiency, placing it as a middle option among three choices.

However, I cannot cite sources because I cannot access the original text data I was trained on. This data is proprietary and confidential, and I cannot share it with anyone. As a result, I cannot provide any specific information about the sources I used to generate my responses."

Surprisingly, Bing AI offered URLs for just 45% of the exam questions, and a significant portion fell into the "very easy" category. However, on a positive note, out of the 35 URLs provided by Bing AI, all were functional, and an impressive 34 were highly relevant to the given context.

## 4. CONCLUSIONS

This research aimed to assess the performance of three platforms incorporating Artificial Intelligence (AI) large language models (LLMs) - ChatGPT, BARD, and Bing AI. These platforms were utilized to solve 20 quantitative questions covering different difficulty levels in an undergraduate finance exam. The exam questions were specifically created for this study, ensuring no access to previous student solutions. The topics covered in the assessment are commonly found in undergraduate finance textbooks.

ChatGPT performed well in addressing "Very Easy" questions with a flawless success rate of 100%. However, it struggled with "Hard" and "Very Hard" questions, scoring 0%. For "Easy" and "Medium" questions, ChatGPT achieved a score of 25%. The model understood the questions' context well and had an 80% accuracy in selecting the appropriate formulas. Nonetheless, limitations included the inability to compute exponents and logarithmic functions and difficulties with iterative computations for IRR and Bond Yield questions.

BARD, on the other hand, scored 15% on the exam. It faced challenges in comprehending question objectives, selecting formulas, and executing computations. BARD often used incorrect formulas and struggled with basic arithmetic operations. Bing AI, running on GPT-4, performed better than BARD but fell short of ChatGPT. It received an overall score of 20% and provided URLs for only 45% of the questions, mostly in the "very easy" category. However, all 35 URLs provided by Bing AI were functional and highly relevant.

All models shared challenges such as inaccurate computations, incorrect formula usage, and difficulty citing sources. Bing AI addressed the citation issue by providing URLs due to its role as a search engine.

In conclusion, ChatGPT showed proficiency in "Very Easy" questions but encountered limitations in higher difficulty levels. BARD struggled in various aspects of the exam, while Bing AI performed better but still below ChatGPT. Improvements are needed to address limitations and minimize the risks of hallucinations. Common challenges include accurate computations and formula selection.

The three AI chatbots assessed in this study are insufficient to assist students in passing the undergraduate finance exam. However, they can serve as valuable tools for dedicated learners. Future technological advancements will likely address the current limitations of AI, enabling improved formula selection and accurate computations involving logarithms, exponents, and more. These advancements can potentially enhance students' performance, enabling them to achieve exceptional scores of 90% or higher in the finance exam.

The paper refrains from offering prescriptions on handling ethical issues in academic integrity but rather presents an overview of generative AI research's current state, limitations in the exam context, and future directions. It emphasizes the need for the academic community to devise reasonable solutions collectively. Given the disruptive nature of generative AI, adaptation becomes imperative for academic community survival and thriving.

## 5. COMPLIANCE WITH ETHICAL STANDARDS

**Funding:** This study is not funded by any entity.

**Conflict of Interest:** Authors do not have any conflicts of interest associated with this article.

**Ethical approval:** This article does not contain any studies with human participants or animals.

**REFERENCES (APA format)**

AlAfnan, M. A., Dishari, S., Jovic, M., & Lomidze, K. (2023). ChatGPT as an Educational Tool: Opportunities, Challenges, and Recommendations for Communication, Business Writing, and Composition Courses. *Journal of Artificial Intelligence and Technology*, *3*(2), Article 2. https://doi.org/10.37965/jait.2023.0184

Berk, J., DeMarzo, P., & Harford, J. (2015). *Fundamentals of Corporate Finance, Global Edition*. The Prentice Hall Ser. in Finance.

Borgaonkar, A. D., Zambrano-Varghese, C. M., Sodhi, J., & Moon, S. (2020, June 22). *Fantastic Cheats: Where and How to Find Them? How to Tackle Them?* 2020 ASEE Virtual Annual Conference Content Access. https://peer.asee.org/fantastic-cheats-where-and-how-to-find-them-how-to-tackle-them

Brealey, R. A., Myers, S. C., Marcus, A. J., Mitra, D., Maynes, E. M., & Lim, W. (2007). *Fundamentals of corporate finance*.

Brigham, E. F., & Houston, J. F. (2021). *Fundamentals of Financial Management: Concise*. Cengage Learning.

Brown, T., Mann, B., Ryder, N., Subbiah, M., Kaplan, J. D., Dhariwal, P., Neelakantan, A., Shyam, P., Sastry, G., Askell, A., Agarwal, S., Herbert-Voss, A., Krueger, G., Henighan, T., Child, R., Ramesh, A., Ziegler, D., Wu, J., Winter, C., … Amodei, D. (2020). Language Models are Few-Shot Learners. *Advances in Neural Information Processing Systems*, *33*, 1877–1901. https://proceedings.neurips.cc/paper/2020/hash/1457c0d6bfcb4967418bfb8ac142f64a-Abstract.html

Dowling, M., & Lucey, B. (2023). ChatGPT for (Finance) research: The Bananarama Conjecture. *Finance Research Letters*, *53*, 103662. https://doi.org/10.1016/j.frl.2023.103662

Drapkin, A. (2023, April 14). Google Bard vs ChatGPT: Which Is the Best AI Chatbot? [Full Test]. *Tech.Co*. https://tech.co/news/google-bard-vs-chatgpt

Dwivedi, Y. K., Kshetri, N., Hughes, L., Slade, E. L., Jeyaraj, A., Kar, A. K., Baabdullah, A. M., Koohang, A., Raghavan, V., Ahuja, M., Albanna, H., Albashrawi, M. A., Al-Busaidi, A. S., Balakrishnan, J., Barlette, Y., Basu, S., Bose, I., Brooks, L., Buhalis, D., … Wright, R. (2023). "So what if ChatGPT wrote it?" Multidisciplinary perspectives on opportunities, challenges and implications of generative conversational AI for research, practice and policy. *International Journal of Information Management*, *71*, 102642. https://doi.org/10.1016/j.ijinfomgt.2023.102642

Hamilton, S. R. (2016, June 26). *Imperative Issues and Elusive Solutions in Academic Integrity: A Case Study*. 2016 ASEE Annual Conference & Exposition. https://peer.asee.org/imperative-issues-and-elusive-solutions-in-academic-integrity-a-case-study

Harding, T. S., Mayhew, M. J., Finelli, C. J., & Carpenter, D. D. (2007). The theory of planned behavior as a model of academic dishonesty in engineering and humanities undergraduates. *Ethics & Behavior*, *17*(3), 255–279.

Hu, L. (2022, November 15). *Generative AI and Future*. Medium. https://pub.towardsai.net/generative-ai-and-future-c3b1695876f2

Jovanović, M., & Campbell, M. (2022). Generative Artificial Intelligence: Trends and Prospects. *Computer*, *55*(10), 107–112. https://doi.org/10.1109/MC.2022.3192720

Jr, B. E. W., & Keith-Spiegel, P. (2001). *Academic Dishonesty: An Educator's Guide*. Psychology Press.

Khan, I. (2023, April 10). *ChatGPT vs. Bing vs. Google Bard: Which AI Is the Most Helpful?* CNET. https://www.cnet.com/tech/services-and-software/chatgpt-vs-bing-vs-google-bard-which-ai-is-the-most-helpful/

Liebrenz, M., Schleifer, R., Buadze, A., Bhugra, D., & Smith, A. (2023). Generating scholarly content with ChatGPT: Ethical challenges for medical publishing. *The Lancet Digital Health*, *5*(3), e105–e106. https://doi.org/10.1016/S2589-7500(23)00019-5

Liu, S., Zappe, S. E., Mena, I. B., Litzinger, T. A., Hochstedt, K. S., & Gallant, T. B. (2015). *Faculty Perspectives About Incorporating Academic Integrity into Engineering Courses*. 26.767.1-26.767.16.

https://peer.asee.org/faculty-perspectives-about-incorporating-academic-integrity-into-engineering-courses

Macfarlane, B., Zhang, J., & Pun, A. (2014). Academic integrity: A review of the literature. *Studies in Higher Education*, *39*(2), 339–358.

OpenAI. (2023). *GPT-4 Technical Report* (arXiv:2303.08774). arXiv. https://doi.org/10.48550/arXiv.2303.08774

Pickering, E., & Schuller, C. (2022). *Widespread usage of Chegg for academic misconduct: Perspective from an audit of Chegg usage within an Australian engineering school*. EdArXiv. https://doi.org/10.35542/osf.io/ds7yb

Radford, A., Narasimhan, K., Salimans, T., & Sutskever, I. (2018). Improving Language Understanding by Generative Pre-Training. *OpenAI*.

Rahimi, F., & Talebi Bezmin Abadi, A. (2023). ChatGPT and Publication Ethics. *Archives of Medical Research*, *54*(3), 272–274. https://doi.org/10.1016/j.arcmed.2023.03.004

Ribas, J. (2023, February 21). *Building the New Bing*. Microsoft Bing Blog. https://blogs.bing.com/search-quality-insights/february-2023/Building-the-New-Bing

Solaiman, I., Brundage, M., Clark, J., Askell, A., Herbert-Voss, A., Wu, J., Radford, A., Krueger, G., Kim, J. W., Kreps, S., McCain, M., Newhouse, A., Blazakis, J., McGuffie, K., & Wang, J. (2019). *Release Strategies and the Social Impacts of Language Models* (arXiv:1908.09203). arXiv. https://doi.org/10.48550/arXiv.1908.09203

Tamkin, A., Brundage, M., Clark, J., & Ganguli, D. (2021, February 4). *Understanding the Capabilities, Limitations, and Societal Impact of Large Language Models*. ArXiv.Org. https://arxiv.org/abs/2102.02503v1

**Appendix (A) – Assessment with Solutions and Explanations**

    Please see the attached file "Appendix A_Assessment."

**Appendix (B) – ChatGPT Results**

    Please see the attached file "Appendix B_ChatGPT."

**Appendix (C) – Bard Results**

    Please see the attached file "Appendix C_Bard."

**Appendix (D) – Bing AI Results**

    Please see the attached file "Appendix D_Bing AI."